\documentclass[12pt,letterpaper]{article}
\pdfoutput=1
\usepackage{wrapfig}
\usepackage{subfigure}
\usepackage{graphicx}
\usepackage{amsfonts}
\usepackage{color}
\newcommand{\be}{\begin{equation}}
\newcommand{\ee}{\end{equation}}
\newcommand{\beq}{\begin{equation}}
\newcommand{\eeq}{\end{equation}}
\newcommand{\bea}{\begin{eqnarray}}
\newcommand{\eea}{\end{eqnarray}}

\def\pd{\partial}

\def\a{\alpha}

\def\s{\sigma}

\def\p{\phi}
\def\te{\theta}
\def\emr{}
\makeatletter \@addtoreset{equation}{section} \makeatother

\begin{document}
%
%
\begin{titlepage}
\hfill MCTP-12-01\\
\begin{center}
{\Large \bf Chaos around Holographic Regge Trajectories}
\vskip .7 cm
\vskip 1 cm
{\large   Pallab Basu,${}^{1}$ Diptarka Das,${}^{1}$ Archisman Ghosh,${}^{1}$}\\
\vskip .4cm
{\large and Leopoldo A. Pando Zayas${}^2$}
\end{center}
\vskip .4cm
\centerline{\it ${}^1$ Department of Physics and Astronomy, University of Kentucky}
\centerline{\it  Lexington, KY 40506, USA}
\vskip .4cm \centerline{\it ${}^2$ Michigan Center for Theoretical
Physics}
\centerline{ \it Randall Laboratory of Physics, The University of
Michigan}
\centerline{\it Ann Arbor, MI 48109-1120}
\vskip .4cm

\begin{abstract}
Using methods of Hamiltonian dynamical systems, we show analytically that a dynamical system connected to the classical spinning string solution holographically dual to the principal Regge trajectory is non-integrable. The Regge trajectories themselves form an integrable island in the total phase space of the dynamical system. Our argument applies to any gravity background dual to confining field theories and we verify it explicitly in various supergravity backgrounds: Klebanov-Strassler, Maldacena-N\'u\~nez, Witten QCD and the AdS soliton. Having established non-integrability for this general class of supergravity backgrounds, we show  explicitly by direct computation of the Poincar\'e sections and the largest Lyapunov exponent, that such strings have chaotic motion.
\end{abstract}
\end{titlepage}

\setcounter{page}{1} \renewcommand{\thefootnote}{\arabic{footnote}}
\setcounter{footnote}{0}
\section{Introduction}

The fact that the quantum numbers of certain operators  or states in field theory can be well described by the corresponding classical string is the idea at the heart of  Regge trajectories where the hadronic relationship $J\sim M^2$ is realized by a spinning string. This fact has a long history dating back to the Chew-Frautschi plots \cite{Chew:1962eu}. In the context of the AdS/CFT correspondence a better understanding of the role played by classical trajectories has been at the center of a substantial part of recent developments. More generally, the AdS/CFT correspondence provides a dictionary that identifies states in string theory with operators in field theory. One of the most prominent examples is provided by the Berenstein-Maldacena-Nastase (BMN) operators. The BMN operators \cite{bmn} can be described as a string moving at the speed of light in the large circle of $S^5$, the operator corresponding to the ground states is given by ${\cal O}^J=(1/\sqrt{JN^J}){\rm Tr}\,\,Z^{J}$. Another interesting 
class of operators which are nicely described as semiclassical strings in the $AdS_5\times S^5$ background are the Gubser-Klebanov-Polyakov (GKP) operators discussed in \cite{gkp}. They are natural generalizations of twist-two operators in QCD and in the context of ${\cal N}=4$ supersymmetric Yang-Mills they look like ${\rm Tr}\Phi^I\nabla_{(a_1} \ldots \nabla_{a_n)}\Phi^I$. A very important property of these operators is that their anomalous dimension can be computed using a simple classical calculation  and yields a prediction for the result at strong coupling $\Delta-S=(\sqrt{\lambda}/\pi)\ln S$. This expression is similar to the QCD relation obtained originally by Gross and Wilczek \cite{gw}.

Right after the original formulation of the AdS/CFT correspondence \cite{Maldacena:1997re,Witten:1998qj,Gubser:1998bc,Aharony:1999ti}  an important direction emerged surrounding the question of how to approach more realistic theories using the methods of the gauge/gravity correspondence. There is by now a well established body of results in this direction. In particular, general conditions on the supergravity backgrounds  have been found that correspond to the existence of the area law for the Wilson loop in the field theory \cite{cobiwl}. Similarly, the classical string configuration corresponding to the Regge trajectories have been extensively studied.

In this paper we study properties of a configuration of classical strings in supergravity backgrounds dual to confining field theories. Our study goes beyond particular trajectories and explores the phase space.  We show that a class of strings that naturally generalizes those corresponding to Regge trajectories is non-integrable. Further, we show explicitly that the motion of such strings is chaotic with the Regge trajectories being an integrable island in the phase space. It turns out that technically the problem is similar to the study of the spectrum of quadratic fluctuations. The study of quantum corrections to Regge trajectories in the context of the AdS/CFT correspondence was initiated in \cite{hep-th/0311190} and was extended to  other backgrounds \cite{hep-th/0409205}. Other recent studies of chaotic behavior of classical strings in the context of the gauge/gravity
correspondence include \cite{arXiv:1007.0277,arXiv:1103.4101,arXiv:1103.4107,arXiv:1105.2540}.  We will in particular draw on modern Hamiltonian methods used in \cite{arXiv:1105.2540} and the concrete discussion of the AdS soliton background presented in \cite{arXiv:1103.4101}.

{\emr
One of the questions driving our program is how to interpret chaos in AdS/CFT, that is, what is the field theory dual of chaotic quantities? We ask whether we can test some of the ideas in the context of confinement. Are there any universal features of various confining theories? We come up with a unified approach to study integrability in a class of confining backgrounds that include many of the commonly-cited examples of confining geometries like Klebanov-Strassler, Maldacena-N\'u\~nez, Witten QCD and AdS-soliton. QCD, in the asymptotic free regime, has been argued to be possibly integrable. One particularly important lead in this direction comes from the integrable Regge trajectories. However our results show that the Regge trajectories are just integrable islands in a wider sea of nonintegrability. One is naturally led to ask the question whether there are more similar subdomains of integrability. In this work, we answer some of the questions above, while some of them still remain open.}

The rest of the paper is organized as follows. In section \ref{sec:spinning} we consider two classes of closed spinning strings and discuss some of their properties in supergravity backgrounds dual to confining field theories.  In section \ref{sec:Analytic}, for the sake of the readers, we review the main results of the literature of analytic non-integrability of Hamiltonian systems. In that section we also show that the motion of the string in supergravity backgrounds dual to confining field theories is non-integrable using analytic methods. Since analytic non-integrability is not a sufficient condition for chaotic behavior, we study numerically a particular background and show strong evidence of chaotic behavior in section \ref{Sec:Chaos}. We conclude in section \ref{Sec:Conclusions}. In  appendix \ref{Sec:Backgrounds} we present the main equation in the non-integrability paradigm of various supergravity backgrounds dual to known confining field theories.

\section{Closed spinning strings in supergravity backgrounds}\label{sec:spinning}
The Polyakov action and the Virasoro constraints characterizing the classical motion of the fundamental string are:
\be
{\cal L}=-\frac{1}{2\pi \alpha'} \sqrt{-g}g^{ab}G_{MN}\partial_a X^M \partial_b X^N,
\ee
where $G_{MN}$ is the spacetime metric of the fixed background, $X^\mu$ are the coordinates of the string, $g_{ab}$ is the worldsheet metric, the indices $a,b$ represent the coordinates on the worldsheet of the string which we denote as $(\tau, \sigma)$. We will use to work in the conformal gauge in which case the Virasoro constraints are
\bea
0&=&G_{MN}\dot{X}^M X'^{N}, \nonumber \\
0&=& G_{MN}\left(\dot{X}^M \dot{X}^N +X'^{M}X'^{N}\right),
\eea
where dot and prime denote derivatives with respect to $\tau$ and $\sigma$ respectively.

We are interested in the classical motion of the strings in background metrics $G_{MN}$ that preserve Poincar\'e invariance
in the coordinates $(X^0,X^i)$ where the dual field theory lives:
\be
ds^2 =a^2(r)dx_\mu dx^\mu + b^2(r)dr^2 +c^2(r) d\Omega_d^2.
\ee
Here $x^\mu=(t,x_1,x_2,x_3)$ and $d\Omega_d^2$ represents the metric on a d-dimensional sub-space that, can also have $r$-dependent coefficients. In the case of supergravity backgrounds in IIB, we have $d=5$ but we leave it arbitrary to also accommodate backgrounds in 11-d supergravity in which case $d=6$.

\newpage
The relevant classical equations of motion for the string sigma model in
this background are
\begin{eqnarray}
\pd_a(a^2(r)\eta^{ab}\pd_b x^\mu)&=&0, \nonumber \\
\pd_a(b^2(r)\eta^{ab}\pd_b r)&=&
\frac{1}{2}\pd_r(a^2(r))\eta^{ab}\partial_ax_\mu\partial_bx^\mu
+\frac{1}{2}\pd_r(b^2(r))\eta^{ab}\partial_ar\partial_br
. \nonumber \\
\end{eqnarray}
They are supplemented by the Virasoro constraints.
We will construct spinning strings by starting with the
following Ansatz ({\bf Ansatz I}):
\begin{eqnarray}
\label{ansatz}
x^0&=&e\, \tau, \nonumber \\
x^1&=&f_1(\tau)\,g_1(\sigma), \qquad x^2=f_2(\tau)\,g_2(\sigma), \nonumber \\
x^3&=& \rm{constant}, \qquad \hspace{0.5cm} r=r(\s).
\end{eqnarray}
We will also consider a slight modification of the above Ansatz as follows ({\bf Ansatz II}):
\begin{eqnarray}
\label{ansatzII}
x^0&=&e\, \tau, \nonumber \\
x^1&=&f_1(\tau)\,g_1(\sigma), \qquad x^2=f_2(\tau)\,g_2(\sigma), \nonumber \\
x^3&=& \rm{constant}, \qquad \hspace{0.5cm} r=r(\tau).
\end{eqnarray}
The main modification is that the radial coordinate is now a function of the worldsheet time $r=r(\tau)$.

With Ansatz I (\ref{ansatz}) the equation of motion for $x^0$ is trivially satisfied.
Let us first show that the form of the functions $f_i$ is
fairly universal for this Ansatz. The equation of motion for $x^i$ is
\be
-a^2\,g_i\ddot f_i + f_i \pd_\s(a^2g_i')=0,
\ee
where a dot denotes a derivative with respect to $\tau$ and
a prime denotes a derivative with respect to $\s$.
Enforcing a natural separation of variables we see that
\be
\label{eq:separation}
\ddot f_i+(e\,\omega)^2 f_i=0, \qquad \pd_\s(a^2\,g_i')+(e\,\omega)^2 a^{2}g_i=0.
\ee
The radial equation of motion is
\be
(b^{2}r')'={1\over 2}\pd_r(a^2)\big[e^2-g_i^2\dot{f}_i^2 + f_i^2 g'{}_i^2\big] + \frac{1}{2}\partial_r(b^2)r'^2.
\ee
Finally the nontrivial Virasoro constraint becomes
\be
b^{2}r'^2+a^{2}\big[-e^2+g_i^2\dot{f}_i^2 + f_i^2 g'{}_i^2\big]=0.
\ee
We are particularly interested in the integrals of motion describing the energy and the angular momentum
\begin{equation}
\label{e}
E={e\over 2\pi \a'}\int a^2  d\s,
\end{equation}
\be
J={1\over 2\pi \a'} \int a^2
\big[x_1\pd_\tau x_2 -x_2\pd_\tau x_1\big] d\s
={1\over 2\pi \a'} \int a^{2}g_1g_2\big[f_1\pd_\tau f_2 -
f_2\pd_\tau f_1 \big] d\s
\ee
The above system can be greatly simplified by further taking the following particular solution:
\be
\label{ansatzB}
f_1=\cos e\omega\,\tau, \quad f_2=\sin e\omega\, \tau, \quad
\rm{and} \quad g_1=g_2=g.
\ee
Under these assumptions the equation of motion for $r$ and the
Virasoro constraint become
\begin{eqnarray}
\label{eom}
&&(b^2r')'-{1\over 2}\pd_r(a^{2})\big[e^2-(e\omega)^2\,g^2 +g'{}^2\big]-\frac{1}{2}\partial_r(b^2)r'^2=0,  \\
&&b^{2}r'{}^2+a^{2}\big[-e^2+(e\omega)^2\, g^2 + g'{}^2\big]=0.
\end{eqnarray}
The angular momentum is then
\begin{equation}
\label{j}
J={e\omega\over 2\pi \a'} \int a^{2}g^2 d\s.
\end{equation}
Since we are
working in Poincar\'e coordinates the quantity canonically conjugate to
time is the energy of the corresponding state in the four dimensional
theory. The angular momentum of the string describes the spin of the
corresponding state. Thus a spinning string in the Poincar\'e coordinates
is dual to a state of energy $E$ and spin $J$. In order for our
semiclassical approximation to be valid we need the value of the action
to be large, this imply that we are considering gauge theory states in
the IR region of the gauge theory with large spin and large energy.
In the cases we study, expressions (\ref{e}) and (\ref{j}) yield a
dispersion relation that can be identified with  Regge
trajectories.
\subsection{Regge trajectories from closed spinning strings in confining backgrounds}
\label{stringconfine}

Let us show that there exists a simple solution of the equations of
motion (\ref{eom}) for any gravity background dual to a confining
gauge theory. The conditions for a SUGRA background to be
dual to a confining theory have been exhaustively explored
\cite{cobiwl} using the fact that the corresponding Wilson loop in field theory should exhibit area law behavior. The main idea is to translate the condition for the vev
of the rectangular Wilson loop to display an area law into properties
that the metric of the supergravity background must satisfy through the identification of the vacuum expectation value of the Wilson loop with the value of the action of the corresponding classical string. It has
been established that one set of necessary conditions is for $g_{00}$
to have a nonzero minimum at some point $r_0$ usually known as the
end of the space wall \cite{cobiwl}.  Note that precisely these two
conditions ensure the existence of  a solution of (\ref{eom}). Namely,
since
$g_{00}=a^{2}$ we see that for a point $r=r_0={\rm constant}$ is a solution if
\begin{equation}
\label{confconds}
\partial_r(g_{00})|_{r=r_0}=0, \qquad g_{00}|_{r=r_0}\ne 0.
\end{equation}
The first condition solves the first equation in (\ref{eom}) and the
second condition makes the second equation nontrivial. Interestingly,
the second condition can be interpreted as enforcing that the
quark-antiquark string tension be nonvanishing as it determines the value of the string action. It is worth mentioning
that due to the UV/IR correspondence
in the gauge/gravity duality the radial direction is  identified with
the energy scale. In particular, $r\approx r_0$ is the gravity
dual of the IR in the gauge theory. Thus, the string we are considering
spins in the region dual to the IR of the gauge theory. Therefore we can
conclude that it
is dual to states in the field theory that are characteristic of the
IR.

Let us now explicitly display the Regge trajectories. The classical
solution is given by (\ref{ansatz}) with  $g(\s)$ solving the second
equation from (\ref{eom}), that is,
 $g(\s)=(1/\omega)\sin(e\omega\s)$. Imposing the periodicity
 $\s\rightarrow \s+ 2\pi$ implies that  $e\omega=1$ and hence
\be
\label{clasreg}
x^0=e\,\tau, \qquad
x^1=e\cos \,\tau \, \sin \,\s,\, \qquad x^2=e\sin \,\tau \, \sin \,\s.
\ee
The expressions
for the energy and angular momentum of the string states are:
\begin{equation}
E=4 { e\, g_{00}(r_0)\over 2\pi \a'}\int d\sigma=2\pi{g_{00}(r_0)} T_s e ,
\qquad J=4{
g_{00}(r_0) e^2\over 2\pi \a'}\int \sin^2\s  d\sigma= \pi{g_{00}(r_0)} T_s e^2 .
\end{equation}

Defining the effective string tension as  ${T_{s,~eff}}=g_{00}(r_0)/(2\pi
\a')$  and $\a'{}_{eff}=\a'/g_{00}$
we find that the  Regge trajectories take the form
\begin{equation}
\label{Regge}
J={1\over 4\pi {T_{s,~eff}}} E^2 \equiv {1\over 2} \alpha'{}_{eff} \,\,t.
\end{equation}
Notice that the main difference with respect to the result in flat space dating back to the hadronic models of the sixties
is that the slope is modified to $\a'{}_{eff}=\a'/g_{00}$.
It is expected that a confining background will have states that
align themselves in Regge  trajectories.

\subsection{Ansatz II}
In this subsection we consider the Ansatz given in equation (\ref{ansatzII}). Note that the analysis given in the previous sections can be applied {\it mutatis mutandis} to this Ansatz. In particular, the separation of variables described in equation (\ref{eq:separation}) can be performed in a symmetric way and one obtains:

\be
g_i'' + \alpha^2 g_i=0, \qquad \partial_\tau(a^2\, \partial_\tau f_i)+\alpha^2 a^2 \, f_i=0.
\ee
The Ansatz  given in (\ref{ansatzII}, \ref{ansatzB}) becomes
\bea
t&=& t(\tau), \qquad r=r(\tau), \nonumber \\
x_1&=& R(\tau)\sin\alpha \sigma, \qquad x_2=R(\tau)\cos\alpha \sigma.
\eea
The Polyakov action is:
\bea
{\cal L}&\propto&a^2(r)\big[-\dot{t}^2 + \dot{R}^2 - \alpha^2 R^2\big] + b^2(r) \dot{r}^2.
\eea
The above Ansatz satisfies the first constraint automatically and the second constraint leads to a Hamiltonian constraint:
\be
a^2(r)[\dot{t}^2+\dot{R}^2+\alpha^2R^2]+b^2(r)\dot{r}^2=0\,.
\ee
We also have that
\be
\label{eq:time_energy}
\dot{t}=E/a^2(r),
\ee
where $E$ is an integration constant. This gives
\be
{\cal L}\propto-\frac{E^2}{a^2(r)}+a^2(r)\big[ \dot{R}^2 - \alpha^2 R^2\big]+ b^2(r) \dot{r}^2\,.
\ee
From the above Lagrangian density the equations of motion for $r(\tau)$ and $R(\tau)$ are
\bea
\label{eqn:main_system}
\frac{d}{d\tau}\left(b^2(r)\frac{d}{d\tau}r(\tau)\right)&=&\frac{{E}^2}{a^3(r)}\frac{d}{dr}a(r) +a(r)\frac{d}{dr}a(r)\big[\dot{R}^2 - \alpha^2 R^2\big]
+b(r)\frac{d}{dr}b(r)(\frac{d}{d\tau}r)^2, \nonumber \\
\frac{d}{d\tau}\left(a^2(r)\frac{d}{d\tau}R(\tau)\right)&=&-\alpha^2 a^2(r)R(\tau).
\eea
We can once again check the claim that for confining backgrounds there is always a confining wall which defines a straight line solution.
Since one can always argue for confining backgrounds,
\be
a(r)\approx a_0-a_2 (r-r_0)^2.
\ee
It is easily seen that in this region both equations above can be satisfied. The equation for $r(\tau)$ is satisfied by $r=r_0$ and $dr/d\tau=0$.
The solution for  $R(\tau)$ is simply
\be
\frac{d^2}{d\tau^2}R(\tau)+\alpha^2R(\tau)=0, \longrightarrow R(\tau)=A\sin(\alpha\tau+ \phi_0).
\ee
This is precisely the solution discussed in the previous section that corresponds to the Regge trajectories in the dual field theory.

\section{Analytic Non-integrability: From Ziglin to Galois Theory}\label{sec:Analytic}
Let us review, for the benefit of the reader, the main statements of the area of analytic
non-integrability  \cite{Fomenko,Morales-Ruiz,Goriely}. First, the term analytic is identified with meromorphic. A meromorphic function on
an open subset D of the complex plane is a function that is holomorphic on all D except a set of isolated points, which are poles of the function. The central place in the study of integrability and non-integrability
of dynamical systems is occupied by ideas developed in the context of the KAM theory. The KAM theorem describes how an integrable
system reacts to small deformations. The loss of integrability is readily characterized by the resonant properties of the
corresponding phase space tori, describing integrals of motion in the action-angle variables. These ideas were already
present in Kovalevskaya's work but were made precise in the context of KAM theory.

Consider a general system of differential equations $\dot{\vec{x}}=\vec{f}(\vec{x})$.
The general basis for proving nonintegrability of such a system  is the
analysis of the variational equation around a particular solution $\bar{x}=\bar{x}(t)$ which is called the straight line solution.  The variational equation
around $\bar{x}(t)$ is a linear system obtained by linearizing the vector field around $\bar{x}(t)$. If the nonlinear
system admits some first integrals so does the variational equation. Thus, proving that the variational equation does
not admit any first integral within a given class of functions implies that the original nonlinear system is
nonintegrable. In particular when one works in the analytic setting where inverting the straight line solution $\bar{x}(t)$, one obtains a (noncompact) Riemann surface $\Gamma$ given by integrating $dt=dw/\dot{\bar{x}}(w)$ with the
appropriate limits. Linearizing the system of differential equations around the straight line solution
yields the {\it Normal Variational Equation } (NVE), which is the component of the linearized system which
describes the variational normal to the surface $\Gamma$.

The methods described here are useful for Hamiltonian systems, luckily for us, the Virasoro constraints in string
theory provide a Hamiltonian for the systems we consider. This is particularly interesting as the origin of this
constraint is strictly stringy but allows a very intuitive interpretation from the dynamical system perspective. One important result at the heart of a analytic non-integrability are Ziglin's theorems. Given
a Hamiltonian system, the main statement of Ziglin's theorems is to relate the existence of a first integral of motion
with the monodromy matrices around the straight line solution \cite{ZiglinI, ZiglinII}. The simplest way to compute
such monodromies is by changing coordinates to bring the normal variational equation into a known
form (hypergeometric, Lam\'e, Bessel, Heun, etc). Basically one needs to compute the monodromies
around the regular singular points. For example, in the case where the NVE is a Gauss hypergeometric
equation $z(1-z)\xi'' +(3/4)(1+z) \xi' +(a/8) \xi=0$, the monodromy matrices can
be expressed in terms of the product of monodromy matrices obtained by taking closed
paths around $z=0$ and $z=1$. In general the answer depends on the parameters of the
equation, that is, on $a$ above. Thus, integrability is reduced to understanding the possible ranges of the parameter $a$.

Morales-Ruiz and Ramis proposed a major improvement on Ziglin's theory by introducing techniques
of differential Galois theory \cite{MR-S,MR-R,MR-R-S}. The key observation is to change the
formulation of integrability from a question of monodromy to a question of the nature of the
Galois group of the NVE. In more classical terms, going back to  Kovalevskaya's formulation, we are interested in
understanding whether the KAM tori are resonant or not. In simpler terms, if their
characteristic frequencies are rational or irrational (see the pedagogical introductions
provided in \cite{Morales-Ruiz,MR-Kovalevskaya}). This statement turns out to be dealt with
most efficiently in terms of the Galois group of the NVE. The key result is now stated as: If the
differential Galois group of the NVE is non-virtually Abelian, that is, the identity connected
component is a non-Abelian group, then the Hamiltonian system is non-integrable. The calculation
of the Galois group is rather intricate, as was the calculation of the monodromies, but the key
simplification comes through the application of Kovacic's algorithm \cite{Kovacic}. Kovacic's algorithm is an
algorithmic implementation of Picard-Vessiot theory (Galois theory applied to linear differential equations) for second order homogeneous linear differential equations with polynomial coefficients and gives a constructive answer to the existence of integrability
by quadratures. Kovacic's algorithm is implemented in most computer algebra software including
Maple and Mathematica. It is a little tedious but straightforward to go through the steps of the
algorithm manually. So, once we write down our NVE in a suitable linear form it becomes a simple
task to check their solvability in quadratures. An important property of the Kovacic's algorithm
is that it works if and only if the system is integrable, thus a failure of completing the algorithm
equates to a proof of non-integrability. This route of declaring systems non-integrable has been
successfully applied to various situations, some interesting examples
include: \cite{Morales-Ramis, Mciejewski,Primitivo, lakatos}. See also \cite{ZiglinABC} for nonintegrability of
generalizations of the H\'enon-Heiles system \cite{MR-Kovalevskaya}. A nice compilation of examples
can be found in  \cite{Morales-Ruiz}. In the context of string theory it was first applied in \cite{arXiv:1105.2540}.

\subsection{Analytic Nonintegrability in Confining Backgrounds}

\subsubsection{Ansatz II}

For confining backgrounds we have that the conditions on $g_{00}$ described in (\ref{confconds}) imply that:
\be
a(r)\approx a_0-a_2 (r-r_0)^2,
\ee
where $a_0$ is the nonzero minimal value of $g_{00}(r_0)$ and the absence of a linear terms indicates that the first derivative at $r_0$ vanishes.

In this region is easy to show that both equations in (\ref{eqn:main_system}) can be satisfied. The equation for $r(\tau)$ is satisfied by $r=r_0$ and $dr/d\tau=0$. The straight line equation for $R(\tau)$ is simply
\be
\frac{d^2}{d\tau^2}R(\tau)+\alpha^2R(\tau)=0, \longrightarrow R(\tau)=A\sin(\alpha\tau+ \phi_0).
\ee

We can now write down the NVE equation by considering an expansion around the {\it straight line} solution, that is,
\be
r=r_0+\eta(\tau).
\ee

We obtain
\be
\label{eq:NVE_general}
\ddot{\eta} +\frac{a_2 {E}^2}{2b_0^2a_0^3}\bigg[1+\frac{2\alpha^2 A^2 a_0^4}{{E}^2}\, \cos 2\alpha \tau\bigg]\eta=0.
\ee
The question of integrability of the system (\ref{eqn:main_system}) has now turned into whether or not the  NVE above can be solved in quadratures. The above equation can be easily recognized as the Mathieu equation. The analysis above has naturally appeared in the context of quantization of Regge trajectories and other classical string configurations. For example, \cite{hep-th/0311190,hep-th/0409205} derived precisely such equation in the study of quantum corrections to the Regge trajectories, those work went on to compute one-loop corrections in both, fermionic and bosonic sectors. Our goal here is different, for us the significance of (\ref{eq:NVE_general}) is as the Normal Variational Equation around the dynamical system (\ref{eqn:main_system}) whose study will inform us about the integrability of the system.

The solution to the above equation (\ref{eq:NVE_general}) in terms of Mathieu functions is
\be
\eta(\tau)=c_1\, \, C(\frac{\theta}{\alpha^2},\frac{\theta\beta}{ 2\alpha^2}, \alpha\, \tau)+
c_2\, \, S(\frac{\theta}{\alpha^2},\frac{\theta\beta}{ 2\alpha^2}, \alpha\, \tau),
\ee
where $c_1$ and $c_2$ are constants and
\be
\theta= \frac{a_2 {E}^2}{2b_0^2a_0^3}, \qquad \beta= \frac{2\alpha^2 A^2 a_0^4}{{E}^2}.
\ee

A beautiful description of a similar situation is presented in \cite{Acosta-Humanez.B} where non-integrability of some Hamiltonians with rational potentials is discussed. In particular, the extended Mathieu equation is considered as an NVE equation
\be
\label{eq:Mathieu}
\ddot{y}=(a+b\sin t+ c\cos t)y.
\ee
Our equation \ref{eq:NVE_general} is of this form with $2\alpha \tau \to t$ and $b=0$. To aid the mathematically minded reader, and to make connection with our introduction to non-integrability in the beginning of section \ref{sec:Analytic}, we show that the extended Mathieu equation can be brought to an algebraic form using $x=e^{it}$ which leads to:

 \be
 y''+\frac{1}{x}y'+\frac{(b+c)x^2+2ax+c-b}{2x^3}y=0.
 \ee
 The above equation is perfectly ameanable to the application of Kovacic's algorithm. It was shown explicitly in \cite{Acosta-Humanez.B} that our case ($b\ne -c$ above) corresponds to a non-integrable equation. More precisely, the Galois group is the connected component of $SL(2,\mathbb{C})$ and the identity component of the Galois group for (\ref{eq:Mathieu}) is exactly $SL(2,\mathbb{C})$, which is a non-Abelian group.

\section{Explicit Chaotic Behavior}\label{Sec:Chaos}
Analytic non-integrability does not, by itself, imply the presence of chaotic behavior. To logically close the circle we should also show chaotic behavior explicitly by computing chaos indicators such as Poincar\'e sections and the largest Lyapunov exponent. Conveniently, the work of some of the authors showed precisely just that. Namely, in \cite{arXiv:1103.4101} it was shown that the spinning string in the AdS soliton supergravity background, which is a background in the class of confining backgrounds we are interested in, is chaotic . Since a separate an exhaustive publication was devoted to strings in the AdS soliton background here we focus in the Maldacena-N\'u\~nez background and show explicitly that non-integrability is accompanied by positive indicators of chaos. We find a rather unifying pictures as both systems behave analogously. Our explicit work provides strong evidence that, indeed, the dynamical system of the classical string which include the Regge trajectory as a particular point in phase 
space is chaotic.

The expression for the functions $a$ and $b$ in the main dynamical system (\ref{eqn:main_system}) can be read directly from the MN background (\ref{eq:MN})(see appendix for details of the background)

\be
a(r)^2=e^{-\phi_0}\frac{\sqrt{\sinh(2r)/2}}{(r\coth 2r -\frac{r^2}{\sinh^2(2r)}-\frac{1}{4})^{1/4}}, \qquad
b(r)^2= \alpha'\, g_s N a(r)^2.
\ee
It is crucial that
\be
\lim_{r\to 0}a(r)^2\to e^{-\phi_0},
\ee
which is a nonzero constant that determines the tension of the confining string.

\subsection{Poincar\'e sections}

An integrable system has the same number of conserved quantities as degrees of freedom.
A convenient way to understand these conserved charges is by looking at the phase space using action-angle variables. Let us assume that we have a system with $N$ position variables $q_i$ with conjugate momenta $p_i$. The phase space is $2N$-dimensional. Integrability means that there are $N$ conserved charges $Q_i=f_i(p,q)$ which are constants of motion. One of them is the energy. These charges define a $N$-dimensional surface in phase space which is a topological torus (KAM torus). The $2N$-dimensional phase space is nicely foliated by these $N$-dimensional tori. In terms of action-angle variables ($I_i,\theta_i$) these tori just become surfaces of constant action.

It is interesting to study what happens to these tori when an integrable Hamiltonian is perturbed by a small nonintegrable piece. The KAM theorem states that most tori survive, but suffer a small deformation \cite{Ott,Hilborn}. However the {\em resonant} tori which have rational ratios of frequencies, i.e. $m_i\omega_i=0$ with $m\in {\cal Q}$, get destroyed and motion on them become chaotic. For small values of the nonintegrable perturbations, these chaotic regions span a very small portion of the phase space and are not readily noticeable in a numerical study. As the strength of the nonintegrable interaction increases, more tori gradually get destroyed. A nicely foliated picture of the phase space is no longer applicable and the trajectories freely explore the entire phase space with energy as the only constraint. In such cases the motion is completely chaotic.

To numerically investigate this gradual disappearance of foliation we look at the Poincar\'e sections. For our system, the phase space has four variables $r,R,p_r,p_R$. If we fix the energy we are in a three dimensional subspace. Now if we start with some initial condition and time-evolve, the motion is confined to a two dimensional torus for the integrable case. This two-dimensional torus intersects the $R=0$ hyperplane at a circle. Taking repeated snapshots of the system as it crosses $R=0$ and plotting the value of $(r,p_r)$, we can reconstruct this circle. Furthermore varying the initial conditions (in particular we set $R(0)=0$, $p_r(0)=0$, vary $r(0)$ and determine $p_R(0)$ from the Virasoro constraint), we can expect to get the foliation structure typical of an integrable system. In the figures below different colors correspond to different values of $r(0)$. Note that for the MN background the confining wall is located at $r_0=0$ and precisely around that point we see islands of integrability.

\begin{figure}[h!]
\centering
\subfigure[$E = 0.316$]{
\includegraphics[scale=0.35]{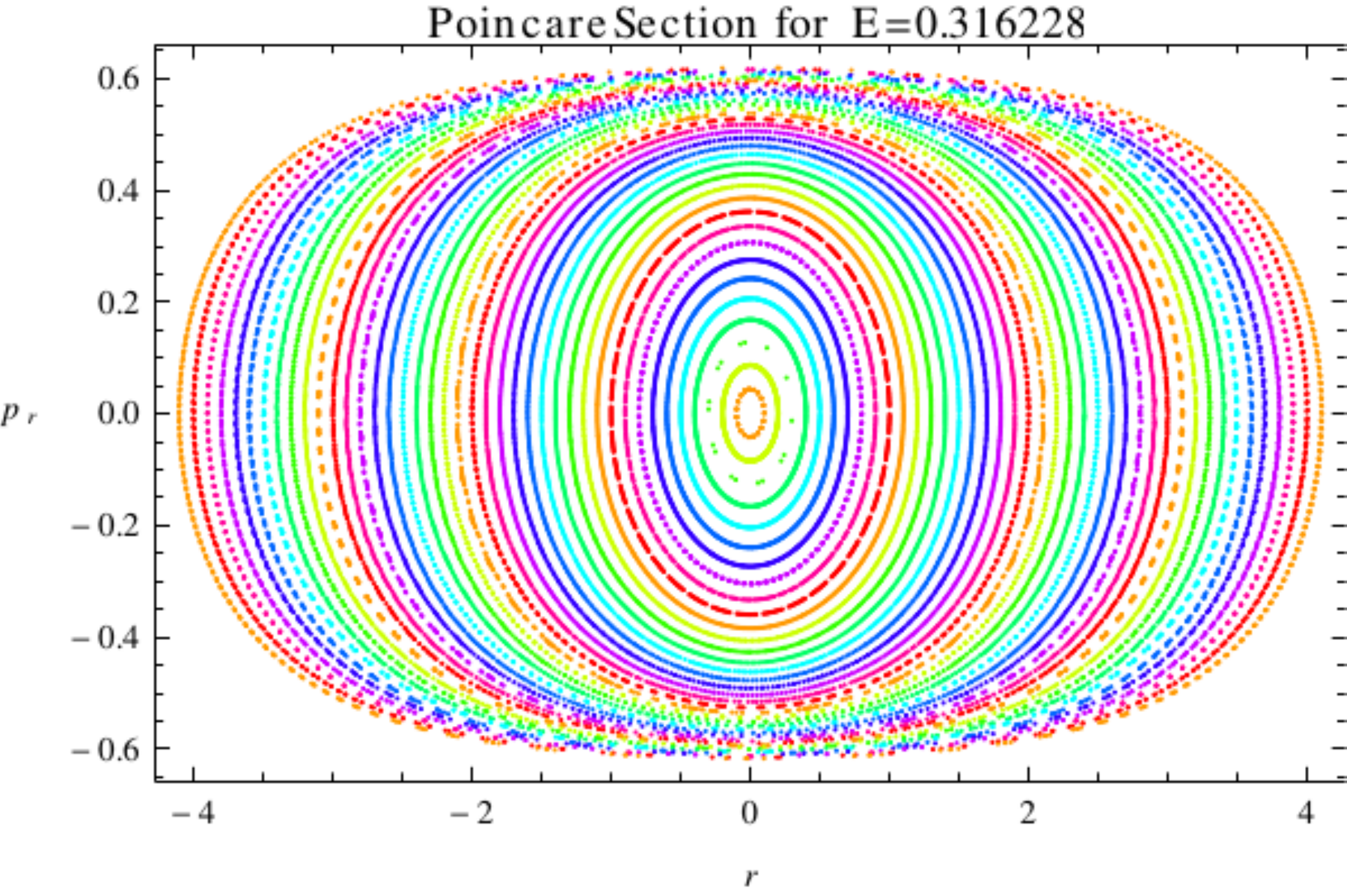}
\label{fig:psec1}
}
\subfigure[$E = 0.5$]{
\includegraphics[scale=0.35]{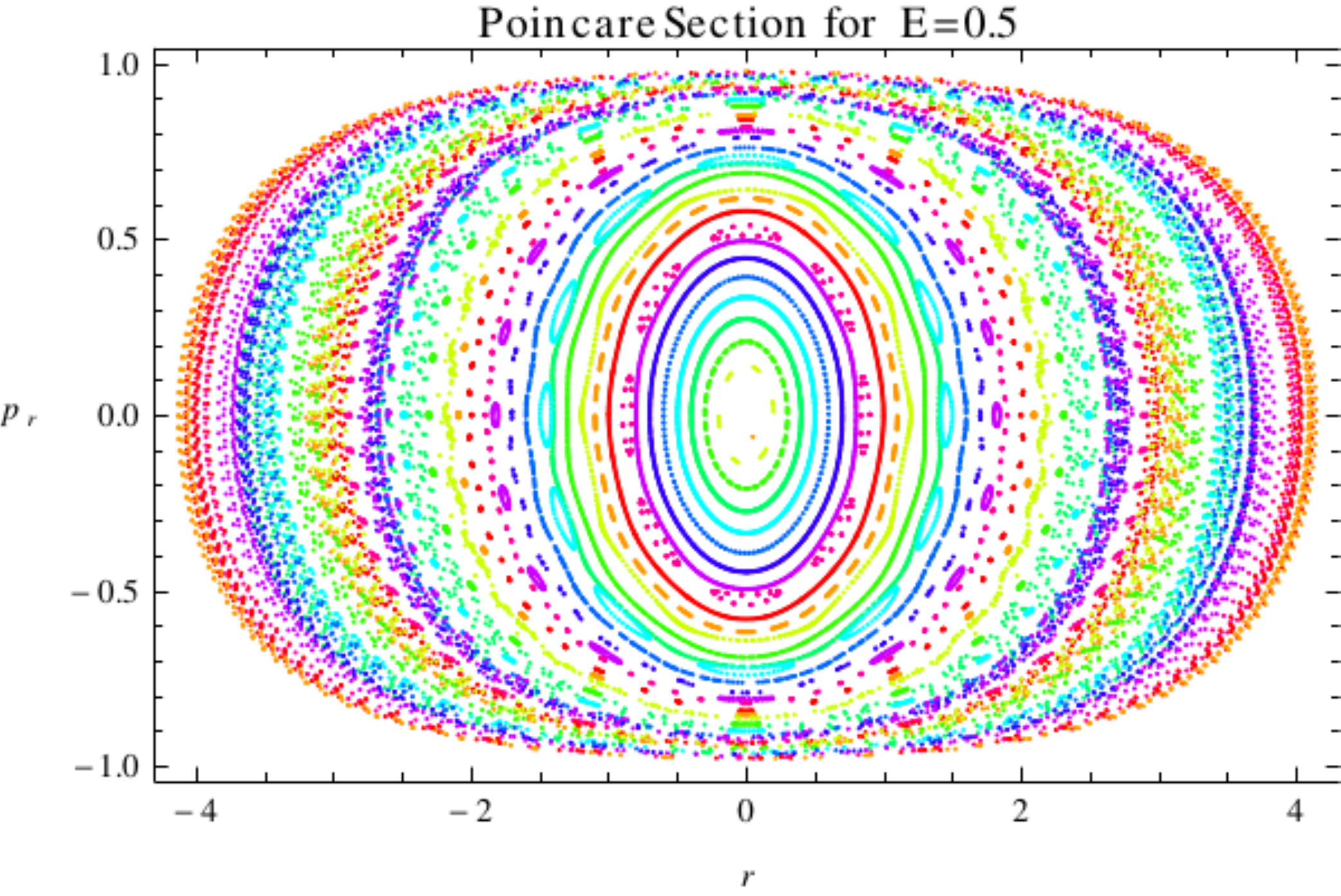}
\label{fig:psec2}
}
\subfigure[$E = 0.71$]{
\includegraphics[scale=0.35]{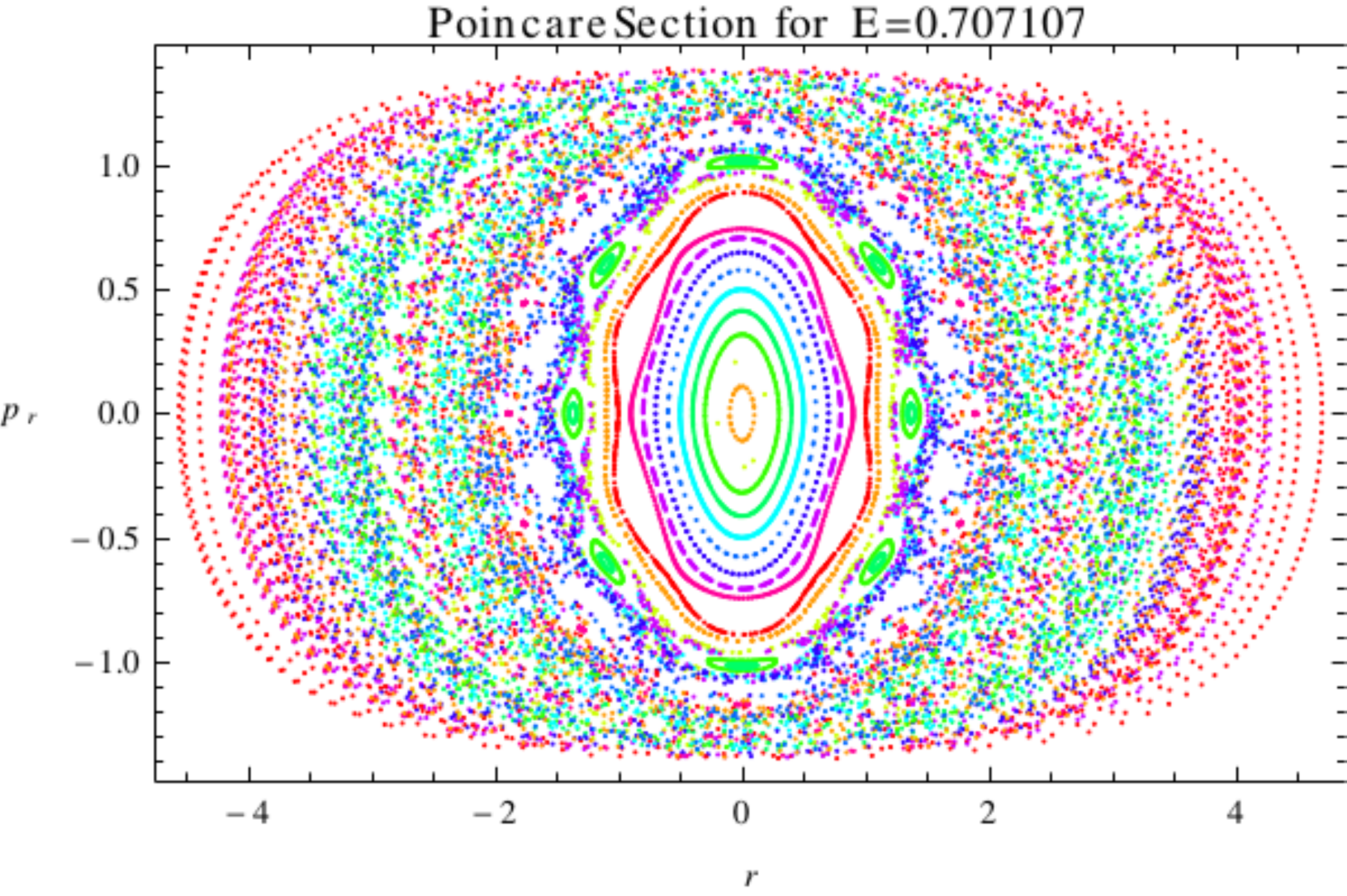}
\label{fig:psec3}
}
\subfigure[$E = 1.0$]{
\includegraphics[scale=0.35]{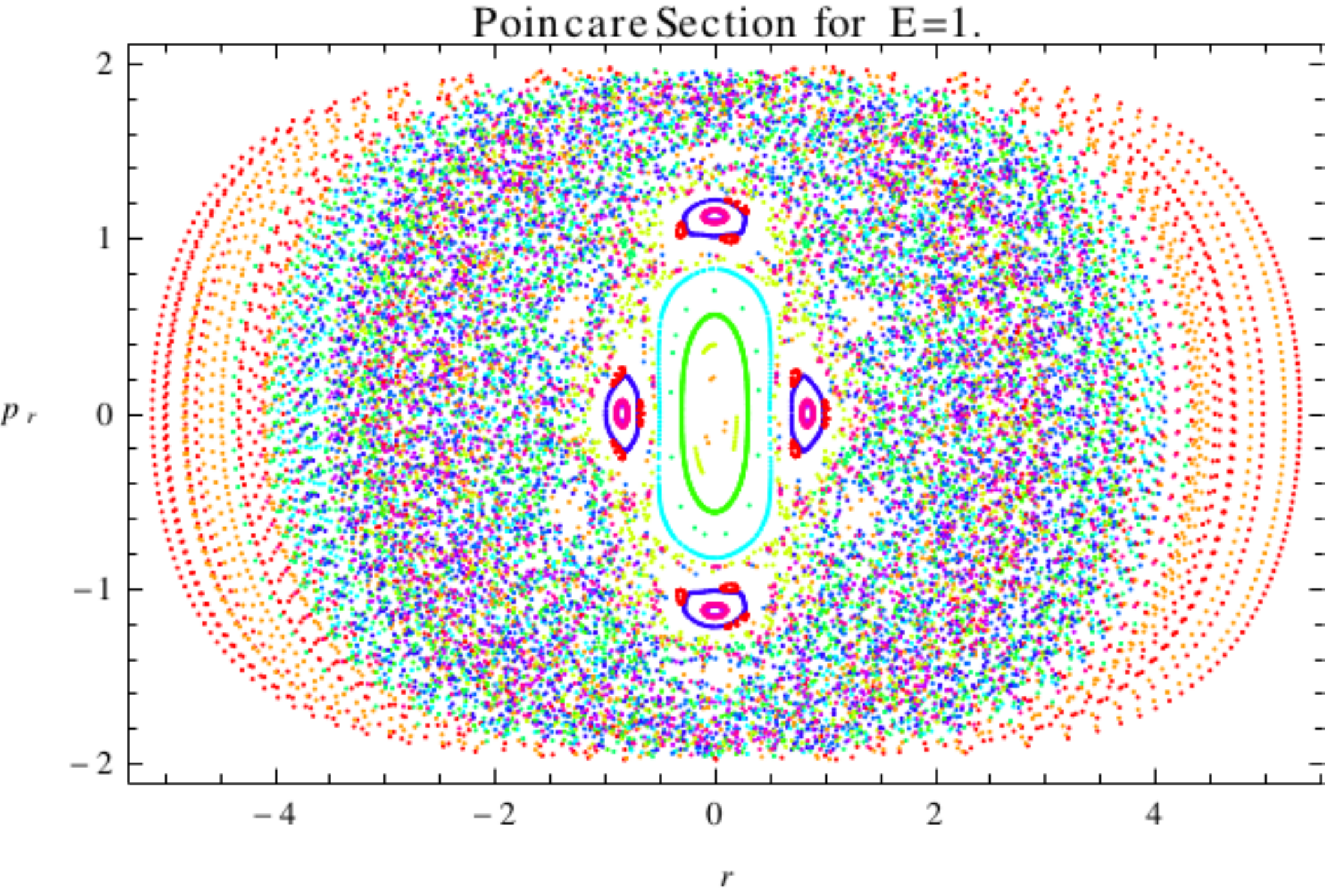}
\label{fig:psec4}
}
\caption{Poincar\'e sections demonstrate breaking of the KAM tori en route to chaos. Each color represents a different initial condition. For smaller values of $E$ the sections of the KAM tori are intact curves, except for the resonant ones. The tori near the resonant ones start breaking as $E$ is increased. For very large values of $E$ all the colors get mixed -- this indicates that all the tori get broken and they fill the entire phase space.}
\label{fig:psec}
\end{figure}

The only parameter in the dynamical system is thus $E$ which we might refer as the energy (being related to the conserved quantity (\ref{eq:time_energy})). Note that this was precisely the case in the analysis of spinning strings in the AdS soliton \cite{arXiv:1103.4101}. Indeed we see that for smaller value of energies ($E$) which is playing the role of the strength of the non-integrable perturbation in the language of KAM theory, a distinct foliation structure exists in the phase space [Fig.\ref{fig:psec1}], as at smaller energy the system may be thought as two decoupled oscillators in $r$ and $R$. Recall that the oscillator with $r_0=0$ corresponds to the Regge trajectory as discussed previously. However as we increase the energy some tori get dissolved [Figs.\ref{fig:psec2},\ref{fig:psec3}.\ref{fig:psec4}]. Although there is no water tight definition of chaos, this destruction of the the KAM tori is one of the strongest indicators of chaotic behavior. The tori which are destroyed sometimes get broken 
down into smaller tori [Figs.\ref{fig:psec2},\ref{fig:psec3}.\ref{fig:psec4}]. Eventually the tori disappear and become a collection of scattered points known as cantori. However the breadths of these cantori are restricted by the undissolved tori and other dynamical elements. Usually they do not span the whole phase space [Figs.\ref{fig:psec4}].    The mechanism is very similar to what happens in well known chaotic systems like H\'enon-Heiles models; our figures are very typical and we refer the reader to the standard text books in this field, for example, \cite{Ott,Hilborn} for qualitative comparison.

\subsection{Lyapunov exponent} \label{subsec:lya}
\begin{figure}[h!]
\centering
\includegraphics[scale=0.70]{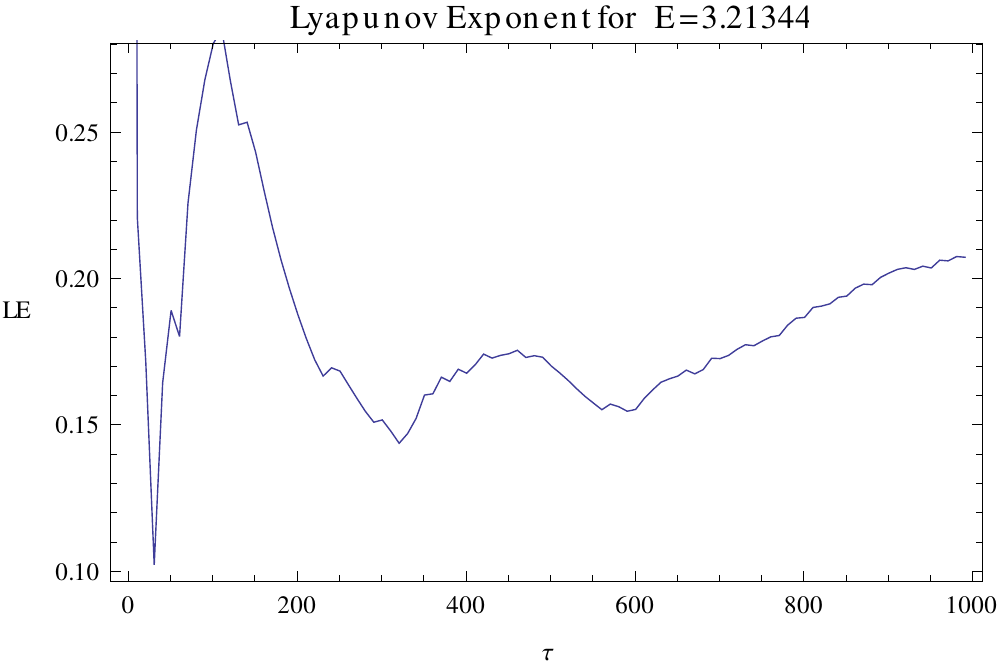}
\label{fig:le}
\caption{The Lyapunov Exponent converges to a positive value of about $0.2$.}
\end{figure}

Let us discuss another important indicator of chaos -- the largest Lyapunov exponent. Sensitivity to the initial conditions is one of the most intuitive characteristics of chaotic systems. More precisely, sensitive dependence on initial conditions means that for some points $X$ in phase space, there is (at least) one point arbitrarily close to $X$ that diverges from $X$. The separation between the two is also a function of the initial location  and has the
form $\Delta X(X_0,\tau)$. The largest Lyapunov exponent is a quantity that characterizes the rate of separation of such infinitesimally close trajectories. Formally it is defined as,
\be
\lambda=\lim_{\tau\rightarrow \infty} \lim_{\Delta X_0 \rightarrow 0} \frac{1}{\tau} \ln \frac{\Delta X(X_0,\tau)}{\Delta X(X_0,0)}
\label{lyapunov}
\ee
In practice we use an algorithm by Sprott \cite{jcsprott}, which calculates $\lambda$ over short intervals and then takes a time average. We should expect to observe that, as time $\tau$ is increased, $\lambda$ settles down to oscillate around a given value. For trajectories belonging to the KAM tori, $\lambda$ is zero, whereas it is expected to be non-zero for a chaotic orbit. We verify such expectations for our case. We calculate $\lambda$ with various initial conditions and parameters. For apparently chaotic orbits we observe a nicely convergent positive $\lambda$ [Fig.\ref{fig:le}].  Our emphasis in not so much in the precise value which might require extensive use of numerical techniques as done in \cite{arXiv:1007.0277}, rather, we are content with showing that the largest Lyapunov exponent is positive.  In figure (\ref{fig:le}) we present a calculation following (\ref{lyapunov}) of the largest Lyapunov exponent. We consider a trajectory with $r(0)=2, R(0)=1.0, p_{r}(0)=0, p_{R}(0)=0$ and its neighbor 
which differs in phase space by $r(0)=r(0)+\epsilon$ with $\epsilon =10^{-3}$.

\section{Conclusions}\label{Sec:Conclusions}
We have established that the motion of certain classical strings in the general class of backgrounds dual to confining theory is chaotic. We have shown analytically, by means of Hamiltonian techniques, that such systems are non-integrable. 
{\emr
One important result of our paper is that non-integrability in confining backgrounds is a direct consequence of the Wilson loop area law. 
The conditions (\ref{confconds}) on the metric, that lead to an area-law behavior of the dual gauge theory thereby implying confinement \cite{cobiwl}, are precisely the conditions required to prove that the string moving in such backgrounds is non-integrable. Non-integrability is thus central to the approach of AdS/CFT to realistic theories.
}

Furthermore, we have also shown numerically that in the case of the MN background the Poincar\'e sections and the largest Lyapunov exponent return positive tests for chaotic behavior.
{\emr Identical results for the AdS soliton background have already been obtained in \cite{arXiv:1103.4101}.  Although, these are the simplest examples in this class of backgrounds that we can explicitly demonstrate to be chaotic, the same result should apply to all theories in the class.}

There are various topics that we find particularly deserving of further attention.  We have established that the classical string trajectory corresponding to the Regge trajectory in field theory is an attractor point in the dynamical system. This same system contains the GKP string which is dual to twist-two operators. It would be interesting to explore in full detail the connection between these two trajectories.

Along similar lines we established in an appendix that Ansatz I can not be chaotic as the effective dynamical ``time'' is periodic. There is {\it a priori} nothing surprising except from the fact that the difference between Ansatz\"e I and II is largely due to $r(\sigma) \to r(\tau)$ which conspicuously looks like a T-duality. This topic is certainly worth exploring.

Lastly, it would be interesting to understand the implication of this classical chaos on the Regge trajectories themselves. Recall that the spectrum of quantum systems obtained as the quantization of systems that in their classical limit are chaotic is quite different from the spectrum of quantum systems obtained from the quantization of integrable classical systems.  This is particularly interesting due to the potential implications for the spectrum of hadronic matter.

\section*{Acknowledgment}
We are thankful to S. Das and A. Shapere for inspiring discussions. We are grateful to Purdue University for organizing a SPOCK meeting and for providing hospitality during the latest stage of the project. The work of L.A.P.Z. is partially supported by Department of Energy under grant DE-FG02-95ER40899 to the University of Michigan. P.B. and A.G. are partially supported by NSF-PHY-0855614. P.B. is also supported by NSF-PHY-0970069.

\appendix
\section {Straight line solution and NVE in Confining Backgrounds }\label{Sec:Backgrounds}
In this appendix we show explicitly that the prototypical supergravity backgrounds in the gauge/gravity correspondence conform to the analysis presented in the main text. We consider the KS and MN
backgrounds explicitly.
\subsection{The Klebanov-Strassler background}
\label{kssection}
We begin by reviewing the KS background, which
is obtained by considering a collection of  $N$  regular
and $M$ fractional D3-branes in the geometry  of the deformed conifold
\cite{ks}.  The 10-d metric is of the form:

\be  ds^2_{10} =   h^{-1/2}(\tau)
dX_\mu dX^\mu  +  h^{1/2}(\tau) ds_6^2
\ ,  \ee  where $ds_6^2$ is the metric of the deformed conifold:

\be
\label{mtmetric}
ds_6^2 = {1\over 2}\varepsilon^{4/3} K(\tau)  \Bigg[ {1\over 3
K^3(\tau)} (d\tau^2 + (g^5)^2)  +  \cosh^2 \left({\tau\over
2}\right) [(g^3)^2 + (g^4)^2]  + \sinh^2 \left({\tau\over
2}\right)  [(g^1)^2 + (g^2)^2] \Bigg].
\ee
where
\be
K(\tau)= { (\sinh (2\tau) -
2\tau)^{1/3}\over 2^{1/3} \sinh \tau},
\ee
and
\bea
\label{forms}
g^1 &=&
{1\over \sqrt{2}}\big[- \sin\theta_1 d\phi_1  -\cos\psi\sin\theta_2
d\phi_2 + \sin\psi d\theta_2\big] ,\nonumber \\  g^2 &=& {1\over
\sqrt{2}}\big[ d\theta_1-  \sin\psi\sin\theta_2 d\phi_2-\cos\psi
d\theta_2\big] , \nonumber \\  g^3 &=& {1\over \sqrt{2}} \big[-
\sin\theta_1 d\phi_1+  \cos\psi\sin\theta_2 d\phi_2-\sin\psi d\theta_2
\big],\nonumber \\  g^4 &=& {1\over \sqrt{2}} \big[ d\theta_1\ +
\sin\psi\sin\theta_2 d\phi_2+\cos\psi d\theta_2\ \big],   \nonumber
\\  g^5 &=& d\psi + \cos\theta_1 d\phi_1+ \cos\theta_2 d\phi_2.
\eea

The warp factor is
given by an integral expression for $h$ is
\be
\label{intsol}
h(\tau) = \alpha { 2^{2/3}\over 4} I(\tau) =  (g_s
M\alpha')^2 2^{2/3} \varepsilon^{-8/3} I(\tau)\ ,
\ee
where
\be
I(\tau) \equiv  \int_\tau^\infty d x {x\coth x-1\over \sinh^2 x}
(\sinh (2x) - 2x)^{1/3}  \ .
\ee
The above integral has the
following expansion in the IR:
\be
I(\tau\to 0) \to a_0 - a_2
\tau^2 + {\cal O}(\tau^4) \ ,
\ee
where $a_0\approx 0.71805$ and
$a_2=2^{2/3}\, 3^{2/3}/18$. The absence  of a linear term in $\tau$
reassures us that we are really expanding  around the end of space,
where the Wilson loop will find it more favorable to arrange itself.
\subsubsection{The straight line solution in KS}
We consider the quadratic fluctuations and their influence on the
Regge trajectories (\ref{Regge}).
In the notation used in the bulk of the paper we have:
\bea
a^2(r)&=&h^{-1/2}(r), \nonumber \\
b^2(r)&=& \frac{\varepsilon^{4/3}}{6K^2(r)}h^{1/2}(r).
\eea
Let us first consider the
metric. The part of the metric perpendicular to the world volume, which is
the deformed conifold metric, does not enter in the classical
solution which involves only world volume fields. Noting that the value
$r_0$ of section \ref{stringconfine} is $\tau=0$, we expand the
deformed conifold up to quadratic terms in the coordinates:
\begin{equation}
\label{defcon}
ds_6^2={\varepsilon^{4/3}\over 2^{2/3} 3^{1/3}}\bigg[{1\over 2} g_5^2
+g_3^2+g_4^2+ {1\over 2} d\tau^2 +{\tau^2\over 4}(g_1^2+g_2^2)\bigg].
\end{equation}
Let us  further discuss the structure of this metric. It is known on
very general grounds that the deformed conifold is a cone over a space
that is topologically $S^3\times S^2$ \cite{candelas}. We can see
that the $S^3$ roughly spanned by $(g_3,g_4,g_5)$ has finite size,
while the $S^2$ spanned by $(g_1,g_2)$ shrinks to zero size at the
apex of the deformed conifold. More importantly for us is the fact that, if we do not allow non-trivial behavior in the directions $(g_1,g_2)$ they cannot contribute to the NVE around the straight line solution characterized by $\tau=0$. Therefore, we have that the NVE equation for the spinning string in the KS background is precisely of the form (\ref{eq:NVE_general}).

\subsection {The Maldacena-N\`u\~nez background}
\label{mnsection}
The MN background \cite{mn} whose IR regime is associated with ${\cal N}=1$
SYM theory is that of a large number of D5 branes wrapping an
$S^2$. To be more precise: (i) the dual field theory to this
SUGRA background is the ${\cal N}=1$ SYM contaminated with KK
modes which cannot be de--coupled from the IR dynamics, (ii) the
IR regime is described by the SUGRA in the vicinity of the origin
where the $S^2$ shrinks to zero size.   The full MN SUGRA background
includes the metric, the  dilaton and the
RR three-form.   It can also be interpreted as uplifting to ten
dimensions a solution of seven dimensional gauged supergravity
\cite{volkov}.  The metric and dilaton of the background are
\bea
\label{eq:MN}
ds^2&=&e^\phi\bigg[dX^adX_a+\a'g_s
N(d\tau^2+e^{2g(\tau)}(e_1^2+e_2^2)+{1\over 4} (e_3^2+e_4^2+e_5^2))\bigg],
\nonumber \\
e^{2\phi}&=&e^{-2\phi_0}{\sinh 2\tau\over 2e^{g(\tau)}}, \nonumber \\
e^{2\,g(\tau)}&=&\tau\coth 2\tau -{\tau^2\over \sinh^2 \, 2\tau}-{1\over
  4}, \nonumber \\
\eea
where,
\bea
e_1&=&d\theta_1, \qquad e_2=\sin\theta_1 d\p_1, \nonumber \\
e_3&=&\cos\psi\, d\te_2+\sin\psi\sin\theta_2\, d\p_2 -a(\tau) d\te_1,
\nonumber \\
e_4&=&-\sin\psi\, d\te_2+\cos\psi\sin\te_2\, d\p_2 -a(\tau) \sin \te_1d\p_1,
\nonumber \\
e_5&=&d\psi +\cos\te_2\, d\p_2 -\cos\te_1d\p_1, \quad
a(\tau)={\tau^2\over \sinh^2\tau}.
\eea
where $\mu=0,1,2,3$, we set the integration constant  $e^{\phi_{D_0}}= \sqrt{g_s N}$.

Note that we use notation where $x^0,x^i$ have dimension of
length whereas $\rho$ and the angles
$\te_1,\phi_1,\te_2,\phi_2,\psi$ are dimensionless and hence the
appearance of the $\alpha'$ in front of the transverse part of
the metric.
\subsubsection{The straight line solution in MN}
The position referred to as $r_0$ in  section (\ref{sec:spinning}) is
$\tau=0$.
Therefore, we will expand the metric around that value. Let us first
identify some structures in the metric that are similar to the deformed
conifold considered in the previous subsection. Notice that $e_1^2+e_2^2$
is precisely an $S^2$. Moreover, near $\tau=0$ we have that
$e^{2g}\approx \tau^2+{\cal O}(\tau^4)$. Thus
$(\tau,e_1,e_2)$ span an $\mathbb{R}^3$  in the limit
\begin{equation}
d\tau^2+e^{2g(\tau)}(e_1^2+e_2^2).
\end{equation}
This means that without exciting the KK modes corresponding to $(e_1,e_2)$ in our Ansatz II, the NVE equation is precisely of the form (\ref{eq:NVE_general}).  Certainly  $e_3^2+e_4^2+e_5^2$ parametrizes a space that is topologically
a three sphere fibered over the $S^2$ spanned by $(e_1,e_2)$. However,
near $\tau=0$ we have a situation very similar to the structure of the
metric in the deformed conifold. Namely, at $\tau=0$ there we have that:
$e_5\to g_5,\,e_3\to \sqrt{2} g_4, \,e_4\to \sqrt{2}g_3$ (up to a
trivial identification $\theta_1\to -\theta_1,\, \phi_1\to -\phi_1$). This allows us
to identify this combination as a round $S^3$ of radius
$2$ and therefore can not alter the form of the NVE (\ref{eq:NVE_general}).


\subsection{The Witten QCD background}
The ten-dimensional string frame metric and
dilaton of the Witten QCD model are given by
\bea
\label{defns}
ds^2&=&({u\over
R})^{3/2} (\eta_{\mu\nu}dx^\mu dx^\nu + {4R^3\over
9u_0}f(u)d\theta^2)+ ({R\over u})^{3/2}{du^2\over f(u)}
+R^{3/2}u^{1/2}d\Omega_4^2\ ,\nonumber \\
f(u)&=&1-{u_0^3\over
u^3}\ , \qquad \qquad \qquad R=(\pi
Ng_s)^{1\over3}{\alpha'}^{1\over2}\ , \nonumber \\
e^\Phi&=&g_s{u^{3/4}\over R^{3/4}}\ .
\eea
The geometry consists of a warped, flat 4-d part, a radial direction $u$, a circle parameterized by $\theta$ with radius vanishing at the horizon $u=u_0$, and a four-sphere whose volume is instead everywhere non-zero.
It is non-singular at $u=u_0$.
Notice that in the $u\to\infty$
limit the dilaton diverges: this implies  that in this limit the
completion of the present IIA model has to be found in M-theory.
The background is completed by a constant four-form field strength
\be
F_4=3R^3\omega_4\ ,
\ee
where $\omega_4$ is the volume form of the
transverse $S^4$.

We will be mainly interested in classical string configurations
localized at the horizon $u=u_0$, since this region is dual to the IR regime of the dual field theory.
In this case the
coordinate $u$ is not suitable because the metric written in
this coordinate looks  singular at $u=u_0$. Then, as a first step, let
us  introduce the radial coordinate
\bea
r^2=\frac{u-u_0}{u_0}\ ,
\eea
so that the metric expanded to quadratic order around $r=0$ becomes
\be
\label{IRmetric}
ds^2\approx ({u_0\over R})^{3/2}[1+{3r^2\over
2}](\eta_{\mu\nu}dx^\mu dx^\nu) + {4\over3}R^{3/2}\sqrt{u_0}(dr^2+ r^2
d\theta^2) + R^{3/2}u_0^{1/2}[1+ {r^2\over 2}]d\Omega_4^2\ .
\ee

\subsubsection{The straight line solution in WQCD}
In this section we consider the closed string configuration corresponding to
the glueball Regge trajectories.
The relevant closed  folded spinning
string configuration dual to the Regge trajectories and constituting the straight line solution in our analysis is

\bea\label{reggesolu} X^0=k\tau\ ,\qquad
X^1=k\cos\tau\sin\sigma\ ,\qquad X^2=k\sin\tau\sin\sigma\ ,
\eea  and
all the other coordinates fixed.

To understand the NVE around the straight line solution given above, we need only look at (\ref{IRmetric}) and realize that the only possible contribution to the NVE given in (\ref{eq:NVE_general}) can come only from KK modes in the $S^4$ of equation (\ref{IRmetric}). We conclude that, in this case, as well the NVE is precisely of the form given in (\ref{eq:NVE_general}).

\section{Comments on Ansatz I}
For confining backgrounds we have that the conditions on $g_{00}$ described in (\ref{confconds}) imply that:
\be
a(r)\approx a_0-a_2 (r-r_0)^2,
\ee
where $a_0$ is the nonzero minimal value of $g_{00}(r_0)$ and the absence of a linear terms indicates that the first derivative at $r_0$ vanishes.

In this region is easy to show that both equations above can be satisfied. The equation for $r(\sigma)$ is satisfied by $r=r_0$ and $dr/d\sigma=0$.
The equation for $R(\sigma)$ is simply
\be
\frac{d^2}{d\sigma^2}R(\sigma)+\omega^2R(\sigma)=0, \longrightarrow R(\sigma)=A\sin(\omega\sigma+ \phi_0).
\ee
We can now write down the NVE equation by considering an expansion around the {\it straight line} solution, that is,
\be
r=r_0+\eta(\sigma).
\ee
We obtain
\be
\label{eq:NVE_generalSigma}
\eta''-\frac{a_2 E^2}{2b_0^2a_0^3}\bigg[1-\frac{2\omega^2 A^2 a_0^4}{E^2}\, \cos 2\omega \sigma\bigg]\eta=0.
\ee
The question of integrability of the system (\ref{eqn:main_system}) has now turned into whether or not the  NVE above can be solved in quadratures. The above equation can be easily recognized as the Mathieu equation. The analysis above has naturally appeared in the context of quantization of Regge trajectories and other classical string configurations. For example, \cite{hep-th/0311190,hep-th/0409205} derived precisely such equation in the study of quantum corrections to the Regge trajectories, those work went on to compute one-loop corrections in both, fermionic and bosonic sectors. Our goal here is different, for us the significance of (\ref{eq:NVE_generalSigma}) is as the Normal Variational Equation around the dynamical system (\ref{eqn:main_system}) whose study will inform us about the integrability of the system.
The general solution to the above equation is
\be
\eta(\sigma)=c_1\, \, C(-\frac{\alpha}{\omega^2},-\frac{\alpha\beta}{ 2\omega^2}, \omega\, \sigma)+
c_2\, \, S(-\frac{\alpha}{\omega^2},-\frac{\alpha\beta}{ 2\omega^2}, \omega\, \sigma),
\ee
where $c_1$ and $c_2$ are constants and
\be
\alpha= \frac{a_2 E^2}{2b_0^2a_0^3}, \qquad \beta= \frac{2\omega^2 A^2 a_0^4}{E^2}.
\ee

Notice, crucially, that although the system obtain here is similar to the one discussed in the main text there is a key difference. Namely, that the effective ``time'' variable $\sigma$ is now periodic. This periodicity precludes us from talking about asymptotic properties which lies at the heart of chaotic behavior. Most indicators of chaos, the largest Lyapunov exponent prominently, are based on the late time asymptotics of the system.


\begin{thebibliography}{99}
\bibitem{Chew:1962eu}
  G.~F.~Chew and S.~C.~Frautschi,
  ``Regge trajectories and the principle of maximum strength for strong interactions,''
  Phys.\ Rev.\ Lett.\  {\bf 8} (1962) 41.

\bibitem{bmn}
D.~Berenstein, J.~M.~Maldacena and H.~Nastase,
``Strings in flat space and pp waves from N = 4 super Yang Mills,''
JHEP {\bf 0204} (2002) 013
[arXiv:hep-th/0202021].

\bibitem{gkp}
S.~S.~Gubser, I.~R.~Klebanov and A.~M.~Polyakov,
``A semi-classical limit of the gauge/string correspondence,''
Nucl.\ Phys.\ B {\bf 636} (2002) 99
[arXiv:hep-th/0204051].


\bibitem{gw}
  D.~J.~Gross and F.~Wilczek,
  Phys.\ Rev.\  D {\bf 8} (1973) 3633.



\bibitem{Maldacena:1997re}
  J.~M.~Maldacena,
  ``The large N limit of superconformal field theories and supergravity,''
  Adv.\ Theor.\ Math.\ Phys.\  {\bf 2} (1998) 231
  [Int.\ J.\ Theor.\ Phys.\  {\bf 38} (1999) 1113]
  [arXiv:hep-th/9711200].
\bibitem{Witten:1998qj}
  E.~Witten,
  ``Anti-de Sitter space and holography,''
  Adv.\ Theor.\ Math.\ Phys.\  {\bf 2} (1998) 253
  [arXiv:hep-th/9802150].
\bibitem{Gubser:1998bc}
  S.~S.~Gubser, I.~R.~Klebanov and A.~M.~Polyakov,
  ``Gauge theory correlators from non-critical string theory,''
  Phys.\ Lett.\  B {\bf 428} (1998) 105
  [arXiv:hep-th/9802109].
\bibitem{Aharony:1999ti}
  O.~Aharony, S.~S.~Gubser, J.~M.~Maldacena, H.~Ooguri and Y.~Oz,
 ``Large N field theories, string theory and gravity,''
  Phys.\ Rept.\  {\bf 323}, 183 (2000)
  [arXiv:hep-th/9905111].

\bibitem{cobiwl}
J.~Sonnenschein,
``Stringy confining Wilson loops,''
arXiv:hep-th/0009146.\\
Y.~Kinar, E.~Schreiber and J.~Sonnenschein, ``Q anti-Q potential
from strings in curved spacetime: Classical results,'' Nucl.\
Phys.\ B {\bf 566} (2000) 103 [arXiv:hep-th/9811192].



\bibitem{hep-th/0311190}
  L.~A.~Pando Zayas, J.~Sonnenschein and D.~Vaman,
  ``Regge trajectories revisited in the gauge / string correspondence,''  Nucl.\ Phys.\ B\ {\bf 682} (2004) 3  [hep-th/0311190].  

\bibitem{hep-th/0409205}
  F.~Bigazzi, A.~L.~Cotrone, L.~Martucci and L.~A.~Pando Zayas,
  ``Wilson loop, Regge trajectory and hadron masses in a Yang-Mills theory from semiclassical strings,''  Phys.\ Rev.\ D\ {\bf 71} (2005) 066002  [hep-th/0409205].  


\bibitem{arXiv:1007.0277}
  L.~A.~Pando Zayas and C.~A.~Terrero-Escalante,
  ``Chaos in the Gauge / Gravity Correspondence,''  JHEP\ {\bf 1009} (2010) 094  [arXiv:1007.0277 [hep-th]].

\bibitem{arXiv:1103.4101}
  P.~Basu, D.~Das and A.~Ghosh,
  ``Integrability Lost,''  Phys.\ Lett.\ B\ {\bf 699} (2011) 388  [arXiv:1103.4101 [hep-th]].

\bibitem{arXiv:1103.4107}
  P.~Basu and L.~A.~Pando Zayas,
 ``Chaos Rules out Integrability of Strings in AdS$_5 \times T^{1,1}$,''  Phys.\ Lett.\ B\ {\bf 700} (2011) 243  [arXiv:1103.4107 [hep-th]].


\bibitem{arXiv:1105.2540}
  P.~Basu and L.~A.~Pando Zayas,
  ``Analytic Non-integrability in String Theory,''  Phys.\ Rev.\ D\ {\bf 84} (2011) 046006  [arXiv:1105.2540 [hep-th]].


\bibitem{Fomenko}
A. T. Fomenko, `` Integrability and Nonintegrability in Geometry and Mechanics,'' Kluwer Academic Publishers, 1988.
\bibitem{Morales-Ruiz}
 Juan Jos\'e Morales Ruiz, ``Differential Galois theory and non-integrability of Hamiltonian Systems, '' Birkh\"auser, Basel 1999.
\bibitem{Goriely}
A. Goriely, ``Integrability and Nonintegrability of Dynamical Systems,'' World Scientific, 2001.
\bibitem{ZiglinI}
S.L. Ziglin, ``Branching of solutions and non-existence of first integrals in
Hamiltonian mechanics I,'' Funct. Anal. Appl. 16 (1982), 181-189.
\bibitem{ZiglinII}
S. L. Ziglin, `` Branching of solutions and non-existence of first integrals in Hamiltonian me-
chanics II,'' Funct. Anal. Appl., 17 (1983), pp. 617.

\bibitem{MR-S}
J.J. Morales-Ruiz and C. Sim\'o, ``Picard-Vessiot theory and Ziglin's Theorem,'' J. Differential Equations 107,  140-162 (1994)
\bibitem{MR-R}
J.J. Morales-Ruiz and J. P. Ramis,``Galoisian obstructions to integrabitly of Hamiltonian Systems I \& II, '' Methods Appl.Anal. 8, 33-111 (2001)
\bibitem{MR-R-S}
J.J. Morales-Ruiz, J.-P. Ramis and C. Sim\`o.  ``Integrability of
Hamiltonian systems and differential Galois groups of higher variational
equations. Ann. Sci. c. Norm. Supr. (4) 40, 845884 (2007)
\bibitem{MR-Kovalevskaya}
J. J. Morales-Ruiz, ``Kovalevskaya, Liapounov, Painleve, Ziglin and
the Differential Galois Theory, '' Regul. Chaotic Dyn. , 2000, 5 (3), 251272.
\bibitem{Kovacic}
J.J. Kovacic, ``An Algorithm for Solving Second Order Linear Homogeneous Differential Equations,'' J. Symb. Comput. {\bf 2} (1986), 3-43.

\bibitem{Morales-Ramis}
J. J. Morales-Ruiz and  J. P. Ramis, `` A Note on the Non-Integrability of Some Hamiltonian Systems with a Homogeneous Potential,''
Methods and Applications of Analysis, c 2001 International Press, Vol. 8, No. 1, pp. 113.120, March 2001.
\bibitem{Mciejewski}
A.J Maciejewski and M. Szydlowski, ``Integrability and Non-Integrability of Planar
Hamiltonian Systems of Cosmological Origin,'' J. Nonl. Math. Phys. 2001, V.8, Supplement, 200206 Proceedings: NEEDS99
\bibitem{Primitivo}
P B. Acosta-Humanez, D. Blazquez-Sanz,
and C. A. Vargas-Contrerar, `` On Hamiltonian Potentials with Quartic Polynomil Normal Variational Equations,''.
\bibitem{lakatos}
R. Lakatos, ``On the Nonintegrability of Hamiltoninan Systems with Two
Degree of Freedom with Homogenous Potential, ''
\bibitem{ZiglinABC}
S.L. Ziglin, ``An Analytic Proof of the Nonintegrability of the ABC-flow for $A=B=C$,'' Funct. Anal. Appl., Volume 37, Number 3, 225-227.


\bibitem{Acosta-Humanez.B}
P. Acosta-Hum\'anez and D. Bl\'azquez-Sanz, ``Non-Integrability of Some Hamiltonians with Rational Potentials,'' Discrete and Continuous Dynamical Systems Serie B, Vol. {\bf 10}, 265-293.


\bibitem{Ott}
E. Ott, ``Chaos in Dynamical Systems,'' Cambridge University Press, Second Edition 2002
\bibitem{Hilborn}
R. Hilborn, ``Chaos and Nonlinear Dynamics: An Introduction for Scientists and Engineers, '' Oxford University Press, Second Edition 2000.
\bibitem{jcsprott}
J. C. Sprott, ``Chaos and Time-Series Analysis,'' Oxford University Press, 2003.


\bibitem{ks}
I.~R.~Klebanov and M.~J.~Strassler,
``Supergravity and a confining gauge theory: Duality cascades
and  chiSB-resolution of naked singularities,''
JHEP {\bf 0008} (2000) 052
[arXiv:hep-th/0007191].

\bibitem{candelas}
P.~Candelas and X.~C.~de la Ossa,
``Comments On Conifolds,''
Nucl.\ Phys.\ B {\bf 342} (1990) 246.

\bibitem{mn}
J.~M.~Maldacena and C.~Nunez,
``Towards the large N limit of pure N = 1 super Yang Mills,''
Phys.\ Rev.\ Lett.\  {\bf 86} (2001) 588
[arXiv:hep-th/0008001].

\bibitem{volkov}
A.~H.~Chamseddine and M.~S.~Volkov,
``Non-Abelian BPS monopoles in N = 4 gauged supergravity,''
Phys.\ Rev.\ Lett.\  {\bf 79} (1997) 3343
[arXiv:hep-th/9707176].

















\end{thebibliography}


\end{document}